\documentclass[journal=achre4,manuscript=article]{achemso}
\pdfoutput=1 
\usepackage[version=3]{mhchem} 
\usepackage{lineno,xcolor}

\usepackage{graphicx}
\usepackage[version=3]{mhchem} 
\usepackage{subfig}
\usepackage{dcolumn}
\newcolumntype{d}[1]{D{.}{\cdot}{#1} }
\usepackage{bm}
\usepackage{siunitx}%
\usepackage{hyperref}

\title{What is moving in hybrid halide perovskite solar cells?}

\author{Jarvist M. Frost}
\author{Aron Walsh$^{\ddag}$}
\email{a.walsh@bath.ac.uk}
\affiliation{Centre for Sustainable Chemical Technologies and Department of Chemistry, University of Bath, Claverton Down, Bath BA2 7AY, United Kingdom}
\affiliation{Global E$^3$ Institute and Department of Materials Science and Engineering, Yonsei University, Seoul 120-749, Korea}

\begin{document}

\newpage

\begin{abstract}
Organic-inorganic semiconductors, which adopt the perovskite crystal structure, have perturbed the landscape of contemporary photovoltaics research.  High-efficiency solar cells can be produced with solution-processed active layers.  The materials are earth abundant and the simple processing required suggests that high-throughput and low-cost manufacture at scale should be possible. 

Whilst these materials bear considerable similarity to traditional inorganic semiconductors, there are notable differences in their optoelectronic behaviour.  A key distinction of these materials is that they are physically soft, leading to considerable thermally activated motion. 

In this Account, we discuss the internal motion of methylammonium lead iodide (\ce{CH3NH3PbI3}) and formamidinium lead iodide (\ce{[CH(NH2)2]PbI3}), covering: (i) molecular rotation-libration in the cuboctahedral cavity; (ii) drift and diffusion of large electron and hole polarons; (iii) transport of charged ionic defects. These processes give rise to a range of properties that are unconventional for photovoltaic materials, including frequency-dependent permittivity, low electron-hole recombination rates, and current-voltage hysteresis. Multi-scale simulations---drawing from electronic structure, \textit{ab initio} molecular dynamic and Monte Carlo computational techniques---have been combined with neutron diffraction measurements, quasi-elastic neutron scattering, and ultra-fast vibrational spectroscopy to qualify the nature and timescales of the motions. Electron and hole motion occurs on a femtosecond timescale. Molecular libration is a sub-picosecond process. Molecular rotations occur with a time constant of several picoseconds depending on the cation. Recent experimental evidence and theoretical models for simultaneous electron and ion transport in these materials has been presented, suggesting they are mixed-mode conductors with similarities to fast-ion conducting metal oxide perovskites developed for battery and fuel cell applications. We expound on the implications of these effects for the photovoltaic action. 

The temporal behaviour displayed by hybrid perovskites introduces a sensitivity in materials characterisation to the time and length scale of the measurement, as well as the history of each sample. It also poses significant challenges for accurate materials modelling and device simulations. There are large differences between the average and local crystal structures, and the nature of charge transport is too complex to be described by common one-dimensional drift-diffusion models. Herein, we critically discuss the atomistic origin of the dynamic processes and the associated chemical disorder intrinsic to crystalline hybrid perovskite semiconductors. 


~

\begin{center}
\includegraphics*{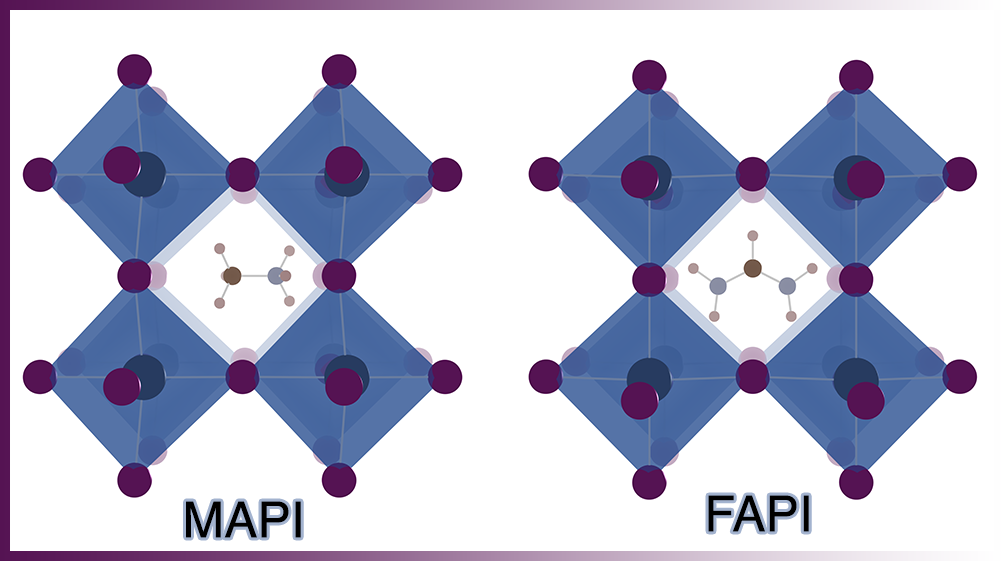}
\end{center}

\end{abstract}



\newpage

\section{Introduction}

Methylammonium lead iodide was first reported in the cubic perovskite crystal structure in 1978.\cite{Weber1978a}
The first solar cell was published in 2009,\cite{Kojima2009} with high-efficiency ($ \geq 10\%$) cells since 2012.\cite{Kim2012a,Lee2012,Liu2013,NamJoongJeon12014,Yang2015d}
There have been many detailed discussions and reviews on the development, progress and limitations of this technology.\cite{Mitzi2001,McGehee2013,Giorgi2014,Butler2014e,DeAngelis2014,Feng2014,Kamat2014b,Frost2014b,Stranks2015b,Even2015b,Padture,Yin2015a,Egger2015a,Walsh2015a}
The focus of this Account is the dynamic behaviour of hybrid organic-inorganic
perovskites, which has been linked to the performance, degradation, and unusual
physics of these materials and their associated devices.
Our principal concern is the atomistic origin of these processes, and their description by first-principle and multi-scale materials modelling.
For brevity, we refer to the methylammonium compound \ce{CH3NH3PbI3} as MAPI and the formamidinium compound \ce{[CH(NH2)2]PbI3} as FAPI.


\section{Molecular Motion}

Weber first reported MAPI in the cubic perovskite crystal structure ($O_h$
symmetry)\cite{Weber1978a}, which may be surprising considering the anisotropy
of the molecular building block (\ce{CH3NH3+} is of $C_{3v}$ symmetry).
However, as early as 1985 it was understood that these molecules were
orientationally disordered in the crystal, giving rise to an effective higher
lattice symmetry on average.\cite{Wasylishen1985}
In 1987, Poglitsch and Weber linked this dynamic disorder to the temperature
and frequency dependence of the complex permittivity.\cite{Poglitsch1987c}
Despite being crystalline solids, the dielectric response of these materials
has one component that is akin to a polar liquid due to the rotational freedom
of the dipolar molecules.\cite{Onoda-yamamuro1992} 
For MAPI at 300 K the contribution of \ce{CH3NH3+} rotations to the static dielectric
response can be estimated as $\sim$ 9, as will be discussed below.

\begin{figure}[]
\begin{center}
\resizebox{12 cm}{!}{\includegraphics*{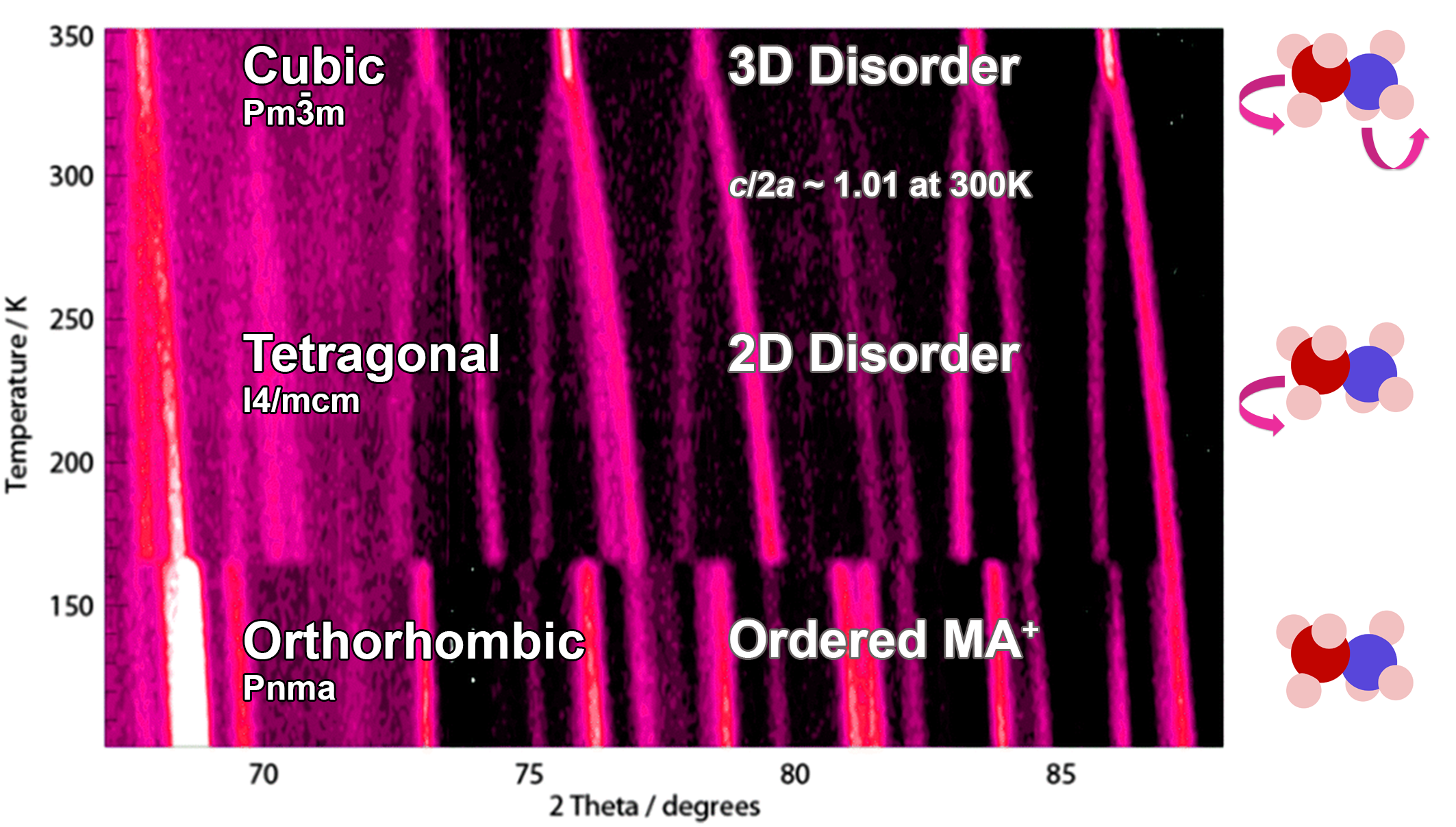}}
\caption{\label{fig-structure} Temperature dependent (100--352 K) powder neutron diffraction
pattern of \ce{CH3NH3PbI3} from Reference ~ \cite{Weller2015a} --- adapted by permission of the Royal Society of Chemistry. 
The space groups of the average crystals structures are shown, along with a schematic of the 
extent of disorder in the \ce{CH3NH3+} sublattice.
} 
\end{center}
\end{figure}

Early work on the characterisation of the crystal structure identified three phases of MAPI: orthorhombic, tetragonal and cubic 
Bravais lattices in order of increasing temperature.\cite{Onoda-yamamuro1992a}
While the position of Bragg peaks in X-ray diffraction can distinguish between the three phases, 
the peak intensities arising from \ce{CH3NH3+} relative to \ce{PbI2-} are too weak to assign accurate molecular orientations.
The recent application of high-resolution powder neutron diffraction (with
a more even distribution of scattering cross-section between light and heavy
atoms) has provided a quantitative description of the temperature dependent average structures (summarised in Figure \ref{fig-structure}).\cite{Weller2015a}
At high temperatures, the structure of MAPI can be represented by the cubic space group $Pm\bar{3}m$, 
whereas below ca. 327 K it is better described by the tetragonal space group $I4/mcm$. 
The cubic-tetragonal transition is second order (Ehrenfest classification) and
continuous over 165--327 K. 
This transition is linked to the anisotropy of the molecular disorder, as well as inorganic octahedral deformation and tilting,
which changes from predominately two dimensions (tetragonal) to three dimensions (cubic). 
The energy balance between the phases can be affected by epitaxial or uniaxial strain.\cite{Ong2015b}
At 165 K, there is a first-order transition to the orthorhombic $Pnma$ phase, with associated discontinuities in 
the observable physical properties.
The transition entropy suggests an ordering of the molecular species,\cite{Onoda-Yamamuro1990}
and a full powder neutron solution shows columnar anti-ferroelectric ordering (head-to-tail arrangement of the \ce{CH3NH3+} cations in 1D channels).\cite{Weller2015a}  
It is possible that other low-temperature orderings can form with different epitaxial
strain, temperature quench regimes, applied electric fields and sample
processing. 

For FAPI, analysis of powder neutron diffraction has assigned a room temperature space group of $Pm\bar{3}m$, which is isostructural to the high-temperature MAPI structure.\cite{Weller2015b}
Faulted or twinned phases at the nanoscale can appear as hexagonal to single-crystal diffraction methods,
which explains the earlier assignment of a hexagonal space group.\cite{Stoumpos2013}
A distinction of FAPI is that the corner-sharing perovskite structure (black colour) is in competition with 
a face-sharing $\delta$ phase (yellow colour);\cite{Stoumpos2013} this is also the case for \ce{CsSnI3}.\cite{Chung2012a,Silva2015}
In contrast to MAPI, knowledge of the low temperature phase behavior of FAPI is currently lacking. 

A point to consider regarding structural characterisation by diffraction, which is particularly relevant to perovskite structured materials, is that standard techniques probe only the average crystal structure.
The intensities of Bragg peaks are related to the average positions of \textit{single} atoms within the unit cell. 
The relative positions of atoms (e.g. bond lengths and angles) requires higher-order correlation functions. 
These can be accessed through using X-ray or neutron total scattering techniques.\cite{Dove2002} 
Disordered crystalline materials display large atomic displacement parameters.
This is the case for both MAPI and FAPI at room temperature.\cite{Weller2015a,Weller2015b}
The material is continuously distorting through this average position as
a function of time --- the thermal vibrations are large relative to the
inter-atomic spacings.

We have attempted to address the issue of molecular motion using three scales of materials modelling: 
(i) static lattice, 
(ii) molecular dynamic and 
(iii) Monte Carlo approaches, 
which we will now address in turn.

\textit{Static Lattice:} the standard approach in density functional theory calculations of solids is to first minimize all forces on the system by structural relaxation. 
In the absence of temperature, the size and shape of the crystallographic unit cell is relaxed with respect to the quantum mechanical forces. 
This should include all internal degrees of freedom.  
It is for this `equilibrium' athermal crystal structure that the electronic structure is calculated, and so the total energy and other properties are extracted. 

As a starting point we assessed the total energy difference of different molecular orientations within a single `quasi-cubic' unit cell of MAPI.\cite{Brivio2013a} 
We note that any electronic structure calculation on such unit cells implicitly
assumes that the methylammonium molecules are arranged into an infinite
ferroelectric domain; although, it should be noted that there is no long-range
electric field created with the standard `tinfoil' boundary
conditions.\cite{tinfoil}
We found that the energy difference between the low index $<100>$ (cube face), $<110>$ (cube edge) and $<111>$ (cube diagonal) orientations of the \ce{CH3NH3+} ions are similar (within 15 meV) with small barriers between them ($<$ 40 meV), but with the facial $<100>$ orientation being preferred. 
Even from this static description, one can envisage molecular orientational disorder driven by thermal energy ($k_BT$ = 26 meV at 300 K).

\textit{Molecular Dynamics:} The most simple method to incorporate temperature into
first-principles simulations is through molecular dynamics (MD). 
The forces on the ions are calculated
quantum-mechanically but the ion velocities are classical and integrated
numerically with Newton's laws of motion to positions. 
Finite temperature is maintained by rescaling ion motion against a thermostat. 
Such simulations should display the full range of anharmonic behaviour, but due
to finite computational resources, rare events are undersampled. 
The stochastic nature of the simulation requires significant post-processing of
data to extract scientific merit. 

Lattice dynamics (phonon calculations) are an alternative but are technically
and computationally more demanding for complex materials such as these. 
This is due to the necessity to find a local potential energy minimum atomic
configuration, and to have sufficiently low noise forces, thereby generating
a well conditioned dynamical matrix. 
At the same time, lattice dynamics only provides information for small
perturbations away from a equilibrium structure. 
The calculated phonons can be directly used to simulate vibrational spectra
within the harmonic approximation.\cite{Brivio2015a,PerezOsorio2015a}

Renderings of our MD simulations\cite{Frost2014} on supercells of MAPI and FAPI are available on-line.\cite{mapi,fapi}  
Without any deep analysis of the trajectories, it is clear to an observer
that the molecules are rotationally mobile at room temperature, with
a timescale of motion on the order of picoseconds.
This motion is consistent with other studies.\cite{Goehry2015a}
A more detailed statistical treatment gives time constants of 3 ps (MA)\cite{Leguy2015b} 
and 2 ps (FA)\cite{Weller2015b} for molecular reorientations at 300 K.

There are a range of complex motions displayed, involving relative twisting of the head and tail groups, libration of a molecular in a single orientation, and the rotation of molecules between orientations.
The results support and expand upon the picture obtained using static lattice techniques:
a statistical analysis\cite{Frost2014} reveals the extent of the disorder as well as the preference for facial $<100>$ orientations that are also evident in neutron diffraction measurements.\cite{Weller2015a,Weller2015b}

A surprising result from the MD simulations is that the \ce{PbI6} octahedra are far from ideal, even for a cubic lattice. 
While the `average' structure may appear cubic the local structure is distorted significantly.
Pb-I-Pb bond angles of 180$^{\circ}$ would be expected for an ideal network, while values of 165 -- 172$^{\circ}$ are found
in the relaxed cubic structure.\cite{data}
Pb(II) is a prototype `lone pair' cation with a ground-state electronic configuration of $5d^{10}6s^{2}6p^{0}$. 
In a centrosymmetric coordination environment, on-site \textit{sp} hybridisation is forbidden by group theory; however, local distortions allow for hybridisation that can result in electronic stabilisation. 
In effect, these systems display a dynamic second-order Jahn-Teller instability,\cite{Walsh2011a,Young2015}
in addition to displacive instabilities associated with rigid titling of the octahedra. 
These displacive motions are particularly important as the electronic bands at the gap
relevant for device operation are hybridised Pb and I atomic orbitals, and will
be directly affected by such distortion. 
These predictions are supported by measurements of the local structure of \ce{CH3NH3SnBr3} 
by X-ray scattering that indicated lone pair distortions.\cite{Worhatch2008}

A limitation with \textit{ab initio} molecular dynamics simulations is that they are restricted to small 
simulation cells (e.g. a $2\times2\times2$ supercell expansion in the majority of our work to date), 
which include a limited set of phonon wavevectors. 
A subtlety of pervoskite molecular dynamics is that the expansions should be
even in the number of octahedra, to avoid falsely constraining the simulation
by preventing zig-zag tilts through the periodic boundaries. 
The development of an analytical interatomic potential model\cite{Mattoni2015a}
can provide access to much larger simulation cells and integration times, and
hence describe temporal correlations over larger areas.

\textit{Monte Carlo:} In order to simulate disordered materials on a scale relevant to photovoltaic devices, we constructed
a model Hamiltonian that describes the intermolecular dipole interactions and
solved for the temperature dependent equilibrium structure using on-lattice Monte Carlo (MC).\cite{Frost2014} 
Since our original report, the \textsc{Starrynight} codes have been generalised to three
dimensional boundary conditions\cite{Leguy2015b} 
and to include defect structures.\cite{Grancini2015a} 
We note that the as-parametrised model considers rotating molecular dipoles, 
but these can be considered an effective dipole of the unit cell including
local lattice distortions, which is also applicable to model fully inorganic perovskites. 
By universality arguments, systems described with the same symmetry and form of
Hamiltonian should have identical transitions and equilibrium behaviour. 

The strength of the methylammonium dipole moment
is comparable to the electric polarisation calculated for simple models of the room temperature tetragonal phase.\cite{Stroppa2015a} 
A molecular dipole of 2.2 Debye $\approx 3\mu$Ccm$^{-2}$ is predicted from quantum chemistry;
by calculating the torque generated around the centre of mass due to an applied electric field, the magnitude of the dipole  is rotation and position invariant even for a charged molecule.
For formamidinium the value is reduced to 0.2 Debye; however, the smaller
molecular dipole can be compensated by larger structural distortions in the solid-state 
owing to the steric effect of the additional amine group.

The general behaviour obtained from the MC simulations is easy to follow: 
ordering of molecular dipoles at low T minimises the total energy of the system 
by maximising the dipole-dipole interactions, while disorder at high T high is driven by configurational entropy.
The electrostatic potential resulting from dipole alignment suggests
that electrons and holes may be segregated by this structure. 
This hypothesis has been subsequently confirmed by
direct electronic structure calculation\cite{Ma2014d} and applied in a device
model.\cite{Sherkar2015}

The molecular motion discussed above has been supported by recent
quasi-elastic neutron scattering (QENS) investigations.\cite{Leguy2015b,chen_rotational_2015}
This technique is particularly sensitive to the motion of hydrogen atoms
due to the high incoherent neutron-scattering cross-section of hydrogen nuclei. 
To get sufficient signal, over 10 g of material was required for these
studies, relying on bulk synthesis procedures which may not be representative
of the active material (thin film) in devices. 
Analysis in one of these studies\cite{chen_rotational_2015} suggests that only ca. 20 \% of the molecules were fully rotationally active in the timescale of the measurement (200 ps), which could be explained by the presence of ordered domains.

Ultra-fast pump-probe vibrational spectroscopy applied to 
device-relevant films offers a direct realspace measurement of methylammonium
motion within MAPI: a fast libration (`wobbling in a cone') motion at 300 fs,
and a slower rotation (reoriention) with a 3 ps signature.\cite{Bakulin2015a}
Here the decay of the polarisation anisotropy within the 10 ps measurement
window indicates that 80 \% of the molecules were rotationally active during
this time, and that any ordered domains must therefore be continuously
interconverting.\cite{Bakulin2015a} 
The time constant for reorientation in FAPI is calculated as 2 ps at 300 K,\cite{Weller2015b}
with no direct measurements available at this time.
The associated frequency range for molecular rotations of 0.3--0.5 THz overlaps
with the low end of the vibrational spectrum, associated with distortions of
the octahedral cage.



Some temperature-dependent device physics is now present in the literature. 
An issue is detangling the temperature-dependence of the active MAPI layer from
the (often organic) contacts. 
There has been one report that below the orthorhombic transition temperature photovoltaic devices stop working abruptly.\cite{Zhang2015h}
More detailed measurements have shown that the photovoltaic power-conversion
efficiency above the orthorhombic transition decreases with decreasing
temperature.\cite{Labram2015a}
They see that the field-effect transistor mobilities decrease with a step
change into the orthorhombic phase.\cite{Labram2015a}  
This is consistent with the model of electrostatic potential domain disorder
being frozen out in the lower temperature phase. 


In order to further understand the device behaviour, we need to understand the
nature and motion of electrical carriers.

\section{Electron and Hole Transport}

There has been significant debate concerning the nature of photo-generated electron-hole pairs in MAPI: free carriers or excitons?\cite{Hirasawa1994a,DInnocenzo2014,Lin2014d,Sheng2015d,Miyata2015}
It is now clear that the small effective mass of the electron and hole
combined with the large static dielectric constant result in long-timescale exciton binding energies, $E_B$
$<< k_bT$ at operating temperatures.\cite{Brivio2014a,Menendez-Proupin2015a,Miyata2015}
The situation can be different in lower dimensional perovskite networks
where exciton binding is enhanced.\cite{Mitzi2001}

The exciton is a quasi-particle---a bound state of an electron polaron and
a hole polaron. 
The long-term energetic stability of these particles is dependent upon the dielectric
function.
The most simple effective mass approximation 
takes the static dielectric constant ($\epsilon_0$), giving a 
binding energy $E_B = \frac{m*}{\epsilon_0^2}$,
where $m*$ represents the reduced carrier mass.
Due to the frequency dependence of this dielectric function (i.e. the range of
response times of the lattice and molecular motion), excitons may be
instantaneously stable yet decay away on the timescales of the dielectric
response with $E_B \sim$ 6 meV at 300 K.\cite{Miyata2015} 
The differing timescales of the dielectric response lead to non-hydrogenic
energy levels of the bound exciton.\cite{Menendez-Proupin2015a}

The calculated high-frequency (optical) dielectric constant of 5,
increases to 24 when the harmonic phonon response of the lattice
is included.\cite{Brivio2014a} 
The additional molecular rotational contribution 
can further increase the static dielectric constant to 33 (assuming an
unhindered dipolar liquid at 300 K).
These values represent a bulk response and exclude any effects from space charges or
conductivity.

An early dielectric study\cite{Onoda-yamamuro1992} shows the strong frequency
dependence (and therefore response time) of the permittivity above 165 
K, which can fitted with a Kirkwood-Fr{\"o}hlich model for a dipolar liquid, 
indicating the increasing hindrance of motion (and thus slower response time,
greater dielectric response) 
of the methylammonium ions upon cooling towards the 165 K transition. 
In the pristine low temperature phase, 
the molecules are fixed and the dielectric response is limited to the electronic and ionic
components with $E_B \sim$ 16 meV.\cite{Miyata2015} 
The dielectric response will have less frequency spread, and more normal
excitonic behaviour will be recovered, with the potential for long-lived
excitons in the orthorhombic perovskite phase. 

%
%
As with any semiconductor, variation in the photophysics and electrical
transport properties reported in the literature can be related to sample
variation 
(stoichiometry, surface, morphology, point and extended defect structures). 
Solution processed materials are particularly susceptible to such variation. 
Care must be taken not to draw general conclusions from a single 
experiment or data source on the hybrid perovskites. 
Some recent reports have used high-quality single crystals whose properties should be more transferable.

%

Hall effect measurements suggest intrinsic dark carrier mobilities of  
\numrange{8}{66}
\si{\centi\metre\squared\per\volt\per\second}
\cite{Stoumpos2013,Wehrenfennig2014d}. 
This is comparable to other solution-processed semiconductors.\cite{Mitzi2004} 
The field-effect mobility is much smaller,   
\numrange{E-2}{E-4} \si{\centi\metre\squared\per\volt\per\second}.\cite{Labram2015a}
These figures are relatively low considering the high dispersion and
well defined bands in the electronic structure of MAPI.
CdTe, which has a similar
band structure to MAPI, displays an electron mobility exceeding \SI{1000}{\centi\metre\squared\per\volt\per\second}
at low temperature.\cite{madelung-04} 
High performance photovoltaics do not directly require high mobility active
materials, rather a sufficient mobility to extract photogenerated charges
before recombination. 

Much more remarkable are the minority-carrier diffusion lengths which exceed \SI{1}{\micro\metre}.\cite{Stranks2013}
This value is large for a direct band gap semiconductor, particularly one which
is solution processed (and therefore structurally defective). 
The value is comparable to high-quality vacuum processed CdTe.
This long carrier diffusion length, being larger than the film thickness 
necessary to absorb the full solar spectrum, makes MAPI and FAPI such high-performance
photovoltaic materials. 

As the carrier mobilities in these materials appear to be modest, the long
diffusion length necessitates an extremely large minority carrier recombination
time---minority-carriers are extremely long lived for an intrinsic (undoped) material. 

From temperature-dependent transport measurements, 
the dominant carrier scattering mechanism has been attributed to acoustic phonons.\cite{Karakus2015,Zhu2015g}
From studying the time constants of transient recombination,  
mono-exponential (suggesting a mono-molecular process) electron-hole recombination is
dominant under normal (sunlight, low fluence) operating conditions, while bi-exponential
recombination (suggesting a bi-molecular process) is a better match to higher light
intensities.\cite{Stranks2015b,Pockett2015a}
The data can be fitted with moderate bulk and interface trap densities, 
but more importantly they suggest the existence of an internal mechanism that
suppresses direct electron-hole interactions.

\begin{figure}[]
\begin{center}
\resizebox{12 cm}{!}{\includegraphics*{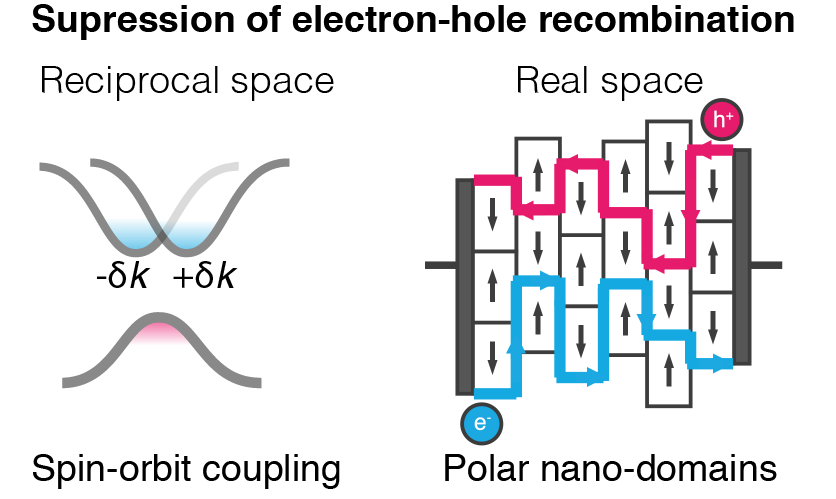}}
\caption{\label{fig-recomb} 
Bi-molecular electron-hole recombination rates in hybrid perovskites are anomalously low. 
Two possible mechanisms are illustrated here: 
(i) relativistic splitting of the band edge states suppresses electron-hole recombination at the valence and conduction band edges due to momentum and spin selection rules.
The band edges are separated in reciprocal space.
(ii) Fluctuations in electrostatic potential caused by the molecular arrangements and/or octahedral distortions could cause electrons and holes to separate (adapted from Ref. \cite {Frost2014b}).
The band edges are separated in real space.
These are two independent models, which could each contribute to a reduced recombination rate.
} 
\end{center}
\end{figure}

There are two likely causes of the non-Langevin recombination in hybrid perovskites:
(i) the local structure of molecular dipoles and octahedra distortions (leading to an inhomogeneity of the electrostatic potential) separates electron and holes in real space;
(ii) the relativistic splitting of the conduction band (a Rashba-Dresselhaus effect\cite{Kepenekian2015a} 
driven by the internal crystal field) results in separation of electrons and holes in reciprocal space.
Polarised domains have been observed from piezoforce microscopy.\cite{Kutes2014a}
Both effects, illustrated in Figure \ref{fig-recomb}, could contribute to the
long minority-carrier lifetimes and deserve more detailed attention from theory, computation and experiment.

A critical question is the spatial extent of the electron and hole carriers in comparison to local molecular order, as this affects both mechanisms
discussed above.
Within the large-polaron theory of Fr\"{o}hlich we previously estimated a coupling constant ($\alpha$) of 1.2.\cite{Frost2014b} 
However, this was based on an estimate for the longitudinal optical phonon frequency of 9 THz, while phonon calculations\cite{Brivio2015a} have revealed a lower lying optical mode at 2.25 THz, which increases $\alpha$ to 2.4.
We can therefore revise our estimate of the effective electron and hole radius
to $\sim$ 4 unit cells and the associated phonon-drag increase in the bare carrier effective mass to 40 \%.



\section{Drift and Diffusion of Charged Ions}


Ion transport in inorganic halide perovskites is well established, with large anion vacancy diffusion coefficients. 
The activation energy in \ce{CsPbCl3} was measured to be 0.29 eV (0.69 eV
including the vacancy formation energy) with a diffusion coefficient of
\SI{2.66E-3}{\centi\metre\squared\per\second}.\cite{MIZUSAKI1983}

In methylammonium lead iodide, we have shown that the Schottky defect formation energy (the energy required to form a stoichiometric amount of isolated charged vacancies) is low (0.14 eV per defect); in  Kr\"oger-Vink notation:
\begin{equation}\label{d1}
\textit{nil} \rightarrow \ce{V_{MA}^/} + \ce{V_{Pb}^{//} + 3\ce{V_I}^{\bullet}} + \ce{MAPbI3}
\end{equation}
Even for a stoichiometric material, there is a thermodynamic driving force for
the formation of 
a high concentration of vacant lattice sites independent of the growth conditions.\cite{Walsh2014b} 
The enthalpic cost of defect formation (0.14 eV per defect) is offset by the gain in configurational entropy of the system
to an extent beyond the dilute defect limit.
Fortunately, these defects do not result in deep electronic states in the band gap,
and thus Shockley-Read-Hall type non-radiative recombination is avoided.\cite{Kim2014b,Yin2015a}

The low formation energy of charged point defects also explains the low carrier concentrations and the difficulty
in extrinsically doping these materials: electronic carriers are heavily compensated by ionic defects, 
which is common in wide band gap semiconductors such as ZnO.\cite{Catlow2011}

A reservoir of available charged point defects in the lattice could support a significant ionic current, as evidenced in impedance spectroscopy of MAPI.\cite{Yang2015e}
However, further to the presence of defects, the mobility of each species depends on the activation energy for solid-state diffusion ($\Delta H^{diff}$), with a hopping rate given by:
\begin{equation}\label{d2}
\Gamma = \nu^* exp\left(-\frac{\Delta H^{diff}}{k_BT}\right)
\end{equation}
where $\nu^*$ represents an effective frequency of the diffusing species in the direction of the saddle point.\cite{Harding1985}

The transport of ions and electrons can be approximated as two separate processes owing to the difference in time constants (see Table \ref{tbl:time}), with the total current density being a sum of the ionic and electron partial current density:
\begin{equation}
j_{total} = j_{ionic} + j_{electronic}
\end{equation}
However, the two processes are invariably coupled. 
Considering a material placed under an applied voltage with ion blocking electrodes, $j_{ionic} = 0$ under steady-state conditions. 
After the current is turned on ions and electrons will flow, but the ions will gradually cease to be available for transport.
The initial equilibration of the ionic distribution relies on the solid-state diffusion pathways discussed above, which are activated 
processes that are slow and can give rise to a temporal change in the current-voltage behaviour before equilibrium is reached.
Electromigration of ions due to momentum transfer from the photovoltaic current in an operating solar cell can also occur,
while ion leakage into the electron and hole contacts is also a possibility.
Ion transport can now be considered the most likely origin of the slow component of the reported hybrid perovskite current-voltage `hysteresis'\cite{Snaith2014} 
and the reversible photocurrents.\cite{Xiao2014b}

\begin{figure}[]
\begin{center}
\resizebox{12 cm}{!}{\includegraphics*{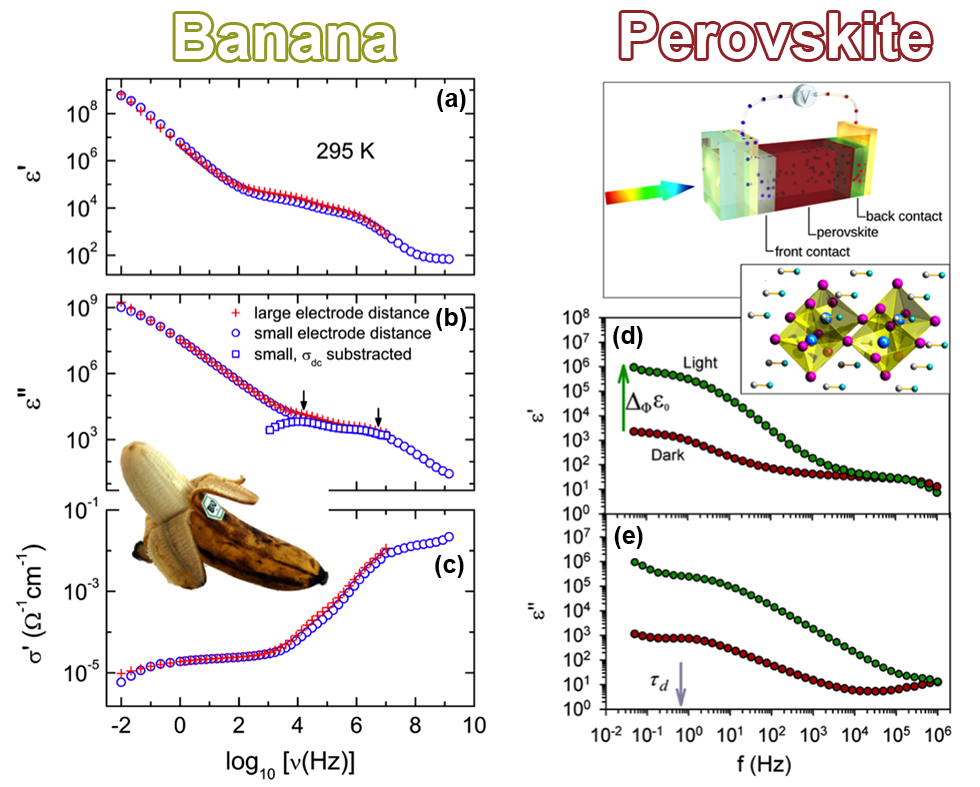}}
\caption{\label{fig-dielectric} 
Comparison of the frequency-dependent dielectric response of a banana (adapted from Ref. \cite{Loidl2008} with permission by the Institute of Physics, Copyright 2008) and \ce{CH3NH3PbI3} (adapted from Ref. \cite{Juarez-Perez2014a} with permission by the American Chemical Society, Copyright 2014).
Broadband spectra of (a) real and (b) imaginary dielectric permittivity and (c) conductivity of a banana skin at room temperature. 
The (d) real and (e) imaginary permittivity of a \ce{CH3NH3PbI3} thin-film under dark and 1 sun illumination conditions.
The free carrier concentration is increased with above bandgap illumination.
} 
\end{center}
\end{figure}

The measured frequency-dependent permittivity of a hybrid perovskite film is compared with the outer-yellow skin of a banana in Figure \ref{fig-dielectric}. 
Both samples show a rapidly increasing dielectric response with decreasing frequency of the applied electric field.
The banana is a good ionic conductor, and an example of a lossy dielectric composite, that shows an apparent hysteresis in current-voltage measurements at low-frequencies in the regime where charging and discharging processes at the electrode interfaces are dominant.\cite{Loidl2008}
The similarity with the response of \ce{CH3NH3PbI3} is notable, with the added complication of contributions from electronic carriers whose concentration are influenced by the light intensity.\cite{Juarez-Perez2014a}

Planar photovoltaic devices will have a build up in electric field (space charge region) 
from ion accumulation
apply evenly across the charge carriers.
In contrast, mesoporous devices will still offer diffusive extraction pathways. 
Rates of ion diffusion will be strongly affected by material stoichiometry
(defect concentration and self-healing ability) and material
composition (diffusion along grain boundaries and interfaces). 
This combination of factors may have led to the large variation in reports of
hysteresis time constants and the `hysteresis-free' devices.\cite{Regan}

The microscopic origin of ion transport is currently controversial with recent proposals of dominant: 
(a) proton diffusion;\cite{Egger2015c}
(b) methylammonium diffusion;\cite{Azpiroz2015a}
(c) iodide diffusion.\cite{Eames2015a,Tateyama}
While the observable response is most likely a combination of several processes, 
the three key factors that determine the contribution of individual ions are
the concentration, activation energy and attempt frequency (Eqn. 2).
Given the structural disorder of these materials discussed previously, the hybrid perovskites are expected to 
exhibit non-ideal diffusion behaviour 
due to the complex kinetics associated with a range of local coordination environments and migration pathways.
Unlike electron transport with electrons and hole wavefunctions delocalised over many unit cells,
ion transport is dependent on and more sensitive to the local structure.

Concerning proton diffusion, free H$^+$ ions are unlikely to present in abundance since methylammonium is a weak acid.
The reaction \ce{CH3NH3+} $\rightleftharpoons$ \ce{CH3NH2} + \ce{H+} has
an equilibrium constant $K_a$ of $\sim 4\times 10^{-11}$ at 300 K.
Taking into account the density of MAPI, this would correspond to a nominal H$^+$ concentration of $10^{11}$cm$^{-3}$, 
which may be affected by the preparatory conditions (e.g. humidity) and reactions with iodine but nonetheless is 
minor compared to the other defects present.

There should be an adequate supply of charged Pb$^{2+}$, I$^-$ and \ce{CH3NH3+}
vacancies with predicted concentrations of $10^{17}-10^{20}$cm$^{-3}$ within the assumption of thermal equilibrium and non-interacting defects.\cite{Walsh2014b}
Due to the orientational disorder of the molecular cations, long-range diffusion of \ce{CH3NH3+} should be slower with transport between successive cages being a relatively complex process, thus lowering the attempt frequency.
The effective frequency should be largest for the iodide ions, as diffusion consists of short jumps over small distances, which is aided by the high polarisability and large thermal displacements for anions in a perovskite lattice.
There is evidence of fast vacancy-mediated anion diffusion in inorganic halide perovskites with a low activation energy, and thus
we consider this to be the dominant process in MAPI and FAPI.
Further evidence is provided by the rapid anion exchange process between the chloride, bromide and iodide perovskites, which  itself is an exceptional phenomenon.\cite{Pellet2015a,Nedelcu2015a}
A cation exchange process has also recently been observed between MAPI and FAPI,\cite{Eperon2015d} which is slower than anion exchange and hence 
consistent with anion diffusion being a faster process. 

\begin{table*}[ht]
\small
 \caption{Summary of dynamic processes and estimates of their associated time constants in \ce{CH3NH3PbI3}.}
  \label{tbl:time}
  \begin{tabular}{lccccc}
    \hline
Process & Microscopic origin & Timescale & Frequency & Diffusion Coefficient \\
    \hline
Lattice vibrations   & Vibrational entropy & 10 fs -- 1 ps  & 1--100 THz  & \\
Molecular libration & Vibrational entropy & 0.5 ps & 2 THz & \\
Molecular rotation  & Rotational entropy & 3 ps & 0.3 THz  & \\
Electron transport  & Drift and diffusion  & $\sim$ 1 fs & $\sim$ 1000 THz & 10$^{-6}$ cm$^2$s$^{-1}$ \\
Hole transport     	& Drift and diffusion  & $\sim$ 1 fs & $\sim$ 1000 THz & 10$^{-6}$ cm$^2$s$^{-1}$ \\
Ion transport   	  	& Drift and diffusion  & $\sim$ 1 ps & $\sim$ 1 THz & 10$^{-12}$ cm$^2$s$^{-1}$\\
    \hline
\end{tabular}
\end{table*}

\section{Conclusions}
The timescales of the various processes discussed in this Account are summarised in Table \ref{tbl:time}.
We now have a reasonable understanding of the isolated motions in this
material. 
There is still much work to do concerning correlated movements, 
including the formation of ordered molecular domains,
the nature of electron and hole polarons,
and the true coupling between electronic and ionic charge carriers.
Our discussion has largely focused on what is moving in \ce{CH3NH3PbI3}, but we have also shown that comparable behaviour is apparent in \ce{[CH(NH2)2]PbI3}.
It is expected that other hybrid perovskites will behave similarly, but there will be differences, e.g. due to different molecular size, shape and/or polarisation.
In particular, the high-performance mixture of 
\ce{(MA)_{1-x}(FA)_xPbI_{3-3x}Br_{3x}} 
could exhibit additional dynamic processes relating to the cation and anion distributions on their respective sub-lattices.
Inorganic perovskites would be expected to exhibit many of these phenomena, with the exception of molecular rotations,
and hence could provide a platform to investigation dynamic disorder with one less degree of freedom.
Whilst halide perovskites may be relatively simple to synthesise, 
they still pose great challenges for a rigorous fundamental understanding.

\begin{acknowledgement}

This work benefited from experimental collaboration with A. Leguy, P. F. Barnes, B. O'Regan, L. M. Peter, P. Cameron, A. A. Bakulin, A. Petrozza and M. T. Weller.
We thank F. Brivio, K. T. Butler and C. H. Hendon, A. B. Walker, M. van Schilfgaarde, M. S. Islam and C. Eames  for useful discussions
and their involvement in the original research discussed here.
The work has been funded by EPSRC Grants EP/K016288/1 and EP/J017361/1, and the ERC (Grant 277757). 
\end{acknowledgement}



\providecommand{\latin}[1]{#1}
\providecommand*\mcitethebibliography{\thebibliography}
\csname @ifundefined\endcsname{endmcitethebibliography}
  {\let\endmcitethebibliography\endthebibliography}{}


\end{document}